# Strains in Axial and Lateral Directions in Carbon Nanotubes

Prashant Jindal[1] and V.K. Jindal[2*]

[1]Department of Mechanical Engineering, Punjab Engineering College
Chandigarh-160012, India

[2]Department of Physics, Panjab University, Chandigarh-160014, India

Analytic expressions for strains in carbon nanotubes of all chiralities in lateral and axial directions have been obtained. These expressions have been related to bond angle and bond length flexibilities. This finally results in obtaining very generalized results for Poisson ratio of all chirality and all diameter carbon nanotubes. It turns out that the Poisson ratio is always independent of carbon nanotube diameter, but depends on the their chirality. On the assumption of fixed bond lengths, (only flexible bond angle), the Poisson ratio of all chirality tubes is obtained to be 1. When realistic flexibility of bond lengths and bond angles is assumed, the Poisson ratio is between ~0 and 0.3, depending upon chirality. The present calculation of the Poisson ratio does not require any detail of molecular interactions under the assumption of either bond angle or bond length flexibility alone, but does depend on the interaction potential if both are assumed flexible. Further, it has been found that the Poisson ratio is sensitive to applied initial stress, varying significantly when obtained under different initial stress values. It is observed that Poisson ratio can be used to justify the interaction potential by finding out the bond length flexibility in comparison to bond angle flexibility.

## I. Introduction

Structural, bulk and dynamical properties of Single Wall (SWNT), Multi-Wall(MWNT) or bunch (CNT bunch) of carbon nanotubes have been extensively studied using either valence-force models, or model potentials and using molecular mechanics simulations.[1-7] In simple computations, based upon assumed interactions, say of Brenner-Tersoff type, one can work out the minimum energy configurations of any of the above CNT formations, to get the equilibrium radius $R_{nm}$ of an assumed initial length of a CNT or a group of CNTs. This configuration can be reworked under stress and equivalently, deviations in the radius coupled to deviations in length can provide the Poisson's ratio, $\nu$. As a result of this method or some alternate methods[1] based upon elastic constants calculations, the theoretical estimates of $\nu$ have been made. In what follows, we show that if bond lengths are assumed rigid in comparison to bond angles, the details of potential are unnecessary for a calculation of $\nu$.

---

[*] The author with whom correspondence be made, e-mail: jindal@pu.ac.in,

The numerical values of various elastic constants as computed using various theoretical approaches differ quite significantly. Generally, Poisson ratio for SWNT have been obtained from the standard elastic constant relations for isotropic linear material,

$$E = 2(1+\upsilon)G, \qquad (1)$$

and

$$E = 3B(1-2\upsilon) \qquad (2)$$

where E, G and B represent Young's modulus, shear modulus and bulk modulus respectively. $\upsilon$ is the Poisson ratio.

The Young's modulus as obtained by several calculations using varying approaches and potentials, as well as obtained experimentally has been well compared by Sears and Batra[1]. Values of Young's modulus computed from experimental data using various techniques, like thermal vibrations, Cantilever bending, tension or Raman spectroscopy are found to differ over a wide range from a value of about 0.27 to 3.6 TPa. The theoretically computed values using various potentials and methods also vary over a wide range, from around 0.66 to 6.8 TPa. In fact, there is quite a variation of understanding about the area of cross section of a SWNT in determining stress or pressure and is one of the causes of such a variation in the value of Young's modulus.

However, a thin sheet (a graphene) folded into a carbon nanotube is just one atom thick. Indeed, in this case, it seems more likely for a redistribution of carbon atoms when compressed in radial dimension to appear elongated in axial direction. In the following we present differences in behaviour of $\upsilon$ for bulk materials, surfaces and caged surfaces. We describe how the axial and lateral strains are related to bond angle and bond length deviations, leading finally to Poisson ratio of SWNT.

## 2.1. Bulk Materials

For a uniform cylinder of radius r, and length l having volume V, we define

$$\pi r^2 l = V, \tag{3}$$

resulting in:

$$\frac{2\Delta r}{r} + \frac{\Delta l}{l} = \frac{\Delta V}{V}, \tag{4}$$

which gives,

$$1 - 2\upsilon = \frac{\Delta V / V}{\Delta l / l}, \tag{5}$$

where the Poisson ratio is defined as:

$$\upsilon = -\frac{\Delta r / r}{\Delta l / l}. \tag{6}$$

Since for isotropic materials, with uniform dilation, an axial stress p gives rise to Young's modulus, E, through $E = p/(\Delta l / l)$, the same axial pressure results in $\Delta V / V = (1/3)p/B$, resulting in Eq. (2). Thus for constant volume under linear strain, $\Delta V / V = 0$, Eq. (5) gives us a value for $\upsilon = 0.5$. If under elongation, the volume reduces (an unlikely situation), $\upsilon \geq 0.5$, otherwise it is smaller than 0.5.

## 2.2. Surfaces

A two dimensional surface having dimensions as l and b and area A, such that $lb = A$, results in

$\frac{\Delta b}{b} + \frac{\Delta l}{l} = \frac{\Delta A}{A}$, thus giving a Poisson ratio as

$$(1 - \upsilon) = \frac{\Delta A / A}{\Delta l / l}. \tag{7}$$

For constant area condition, $\upsilon = 1$.

Again, from uniform area expansion under length expansion to constant area, the value of $\upsilon$ would range from $-1$ to 1 in this case. If we assume a fixed thickness of the surface as t, the ratio of area strains to length strains can be written as E/2B, returning us a new Eq. instead of 2.

## 3. Graphene Sheets and Carbon Nanotubes

The bond angles in the planner sheet of regular hexagons formed by bond lengths *a* are $2\pi/3$ rad and a carbon nanotube of a given chirality is obtained by rolling a graphene sheet (a typical graphene sheet is shown in Fig.1.) along a vector *r* defined in terms of basic vectors $\mathbf{b}_1$ and $\mathbf{b}_2$. A pair of integers, (n,m) defines a vector, called a chiral vector which determines characteristics of a CNT,

$$\mathbf{r} = n\mathbf{b}_1 + m\mathbf{b}_2 .\tag{8}$$

This chiral vector approximately equals the circumference of a nanotube of radius $R_{nm}$.

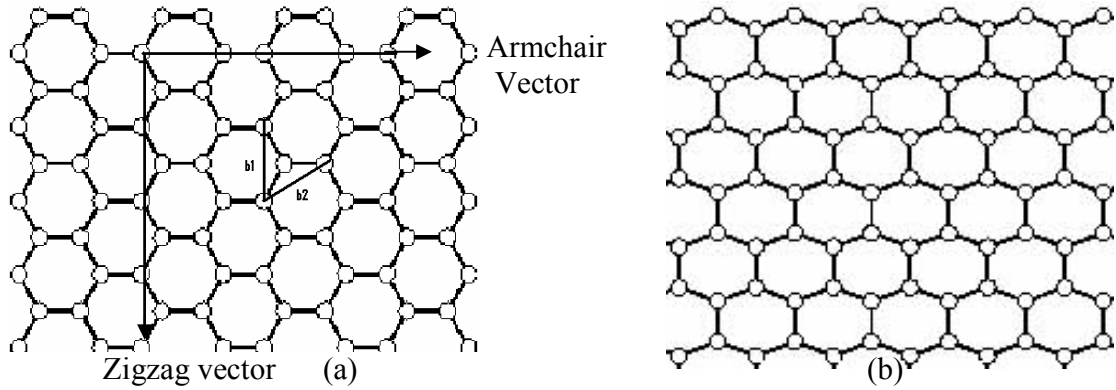

**Fig. 1.** A typical grapheme sheet, showing Armchair and Zigzag Chiral Vectors. The basic vectors defining a chiral tube are shown as $\mathbf{b}_1$ and $\mathbf{b}_2$. Right side figure (b) shows a compressed zigzag CNT, cut and spread to form a sheet.

Assuming that the bond lengths of a regular hexagon (*a*) remain unaltered under a uniform stress perpendicular to a bond, the regular hexagons transform as shown in Fig.2. In this case, the structures of nanotubes are formed by hexagons having bond lengths as *a*, bond angles as $\theta$ (shown in figs. as smaller angle) and $\pi - \theta/2$ (larger angle). It is important to note that we define here the bond angle by $\theta$, an equivalent way to define the same could be the angle $\theta' = \pi - \theta/2$.

The angle $\gamma$ between the basic vectors $\mathbf{b}_1$ and $\mathbf{b}_2$ (of length b) and b are given by:

$$b = 2a\cos\theta/4 ;\tag{9a}$$

$$\gamma = \theta/2 .\tag{9b}$$

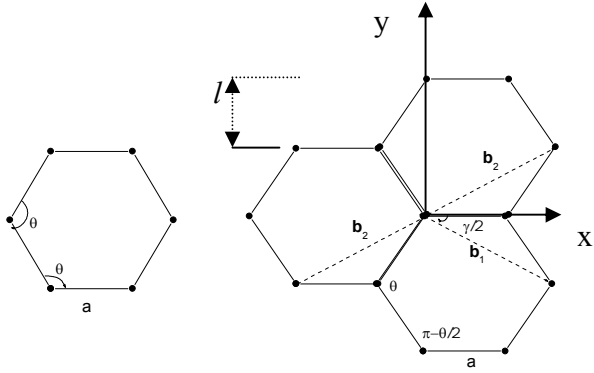

Fig. 2. Unstressed (regular) hexagon and stressed set of hexagons. The stressed configuration assumes bond lengths to be unchanged, hexagon angles get altered to a set of unequal angles as shown.

As a result, chiral vector (Eq. 8) defines the circumference of a (n,m) CNT having radius $R_{nm}$,

$$R_{nm} = |n\mathbf{b}_1 + m\mathbf{b}_2|/2\pi. \qquad (10)$$

Expressing basic vectors in Cartesian coordinates (Fig. 2), the chiral vector can also be written as:

$$\mathbf{r}_{nm} = 2a\cos(\theta/4)\left[(n+m)\cos(\theta/4)\mathbf{x} - (n-m)\sin(\theta/4)\mathbf{y}\right]. \qquad (11)$$

This helps in identifying a vector in a perpendicular direction, which is along the length of the CNT. It is given as

$$\mathbf{l}_{nm} = c\left[(n-m)\sin(\theta/4)\mathbf{x} + (n+m)\cos(\theta/4)\mathbf{y}\right], \qquad (12)$$

where c is a multiplying constant which depends on any arbitrary length of the CNT. Eq. (10) leads to

$$R_{nm} = \frac{2a\cos(\theta/4)}{2\pi}\rho_{nm}, \text{ with } \rho_{nm} = \sqrt{n^2 + m^2 + 2nm\cos(\theta/2)}. \qquad (13)$$

This equation relates the radius of any (n,m) CNT, with a bond length *a* and any bond angle $\theta$, which can be different from $2\pi/3$ under stress. For $\theta = 120°$, these reduce to results for unstressed CNT available in literature.

Fractional transverse strain $\Delta R/R$ is easily obtained from Eq. (13) and given by:

$$-\frac{\Delta R}{R} = \left[\frac{1}{4}\tan(\theta/4) + \frac{1}{2}\frac{nm\sin(\theta/2)}{n^2 + m^2 + 2nm\cos(\theta/2)}\right]d\theta - \frac{\Delta a}{a}. \quad (14)$$

At zero stress, the bond angle is close to $2\pi/3$. Therefore, at zero stress we write

$$\left(\frac{\Delta R}{R}\right)_0 = -[0.144 + 0.433I/(1 + I^2 + I)]d\theta + \frac{\Delta a}{a} \quad (15)$$

where $I$ defines the chirality, $I=n/m$.

Here $\frac{\Delta a}{a}$ is the fractional change in the bond length which results from stress applied on a given tube, in addition to the change in the bond angle $\theta$ by $d\theta$. We assume that the bond length is altered uniformly and each bond of the hexagon alters by the same amount.

A minimum length of the CNT having repetitive character is given by the perpendicular distance $l$ between two parallel chiral vectors, e.g., $\mathbf{r} = n\mathbf{b}_1 + m\mathbf{b}_2$ and $\mathbf{r}'$ as shown in Fig.4.

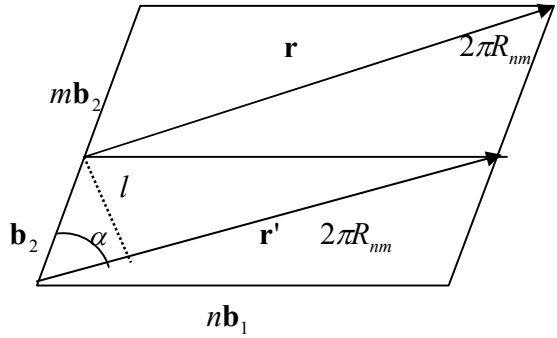

**Fig. 4** Diagram showing the minimum equivalent length of a (n,m) carbon nanotube.

The length $l$ of a general chiral nanotube is given by:

$$l = b\sin\alpha = |\mathbf{b}_2 \times \mathbf{r}_{nm}|/r_{nm} = n|\mathbf{b}_2 \times \mathbf{b}_1|/(2\pi R_{nm}), \quad (16)$$

which finally gives

$$l = \frac{4na^2 \cos^2(\theta/4)\sin(\theta/2)}{2\pi R_{nm}}. \quad (17)$$

Therefore, assuming that both the bond length and the bond angle are responsible for the length strain, $\Delta l / l$, which is independent of the total length of the tube chosen (the longitudinal strain) can be easily obtained in the form,

$$\frac{\Delta l}{l} = \left[\frac{1}{2}\{\cot(\theta/2) - \tan(\theta/4)\} - \left(\frac{\Delta R}{R}\right)_a\right] d\theta + \frac{\Delta a}{a}. \tag{18}$$

Where $\left(\dfrac{\Delta R}{R}\right)_a$ is the fractional radial strain at constant bond length. It is given by the first part of Eq. (14).

Again, at zero stress, Eq. 17 reduces to:

$$\left(\frac{\Delta l}{l}\right)_0 = [0.144 + 0.433 I(1 + I^2 + I)] d\theta + \frac{\Delta a}{a} \tag{19}$$

And for Poisson ratio of unstressed CNTs as

$$\nu = \frac{\alpha - \kappa}{\alpha + \kappa}, \quad \kappa = \frac{1}{a}\frac{\Delta a}{\Delta \theta} \tag{20}$$

where $\alpha = 0.144 + 0.433 I(1 + I^2 + I)$.

For armchair and zigzag nanotubes, we can obtain the results directly.

The numerical results for lateral strain, axial strain and Poisson ratio using Eqs. 15, 19 and 20 are easily obtained. These are presented in Fig. 5 for various values of the bond length variation with respect to bond angle.

## 4. Special Cases

For armchair CNT, n=m, I=1. As a result, we obtain, $\alpha_{nn} = 0.289$. Similarly, for zigzag tubes, I=0, and we obtain $\alpha_{n0} = 0.144$.

It is thus easily seen that deviation in fractional bond length ($\kappa$) can have only up to 28.9% value of the deviation in bond angle for a given applied stress in order for us to have a positive value for the Poisson ratio for armchair tubes. It reduces to 14.4% in case of zigzag tubes.

A realistic estimate of the fractional bond length variation with respect to bond angle can be picked up from various calculations done using model potentials and MD calculations where the bond lengths and bond angles are plotted at various stress values.

From such data[6,7], it seems close to about 20%. The results presented in Fig. 5c show Poisson ratio using this data.

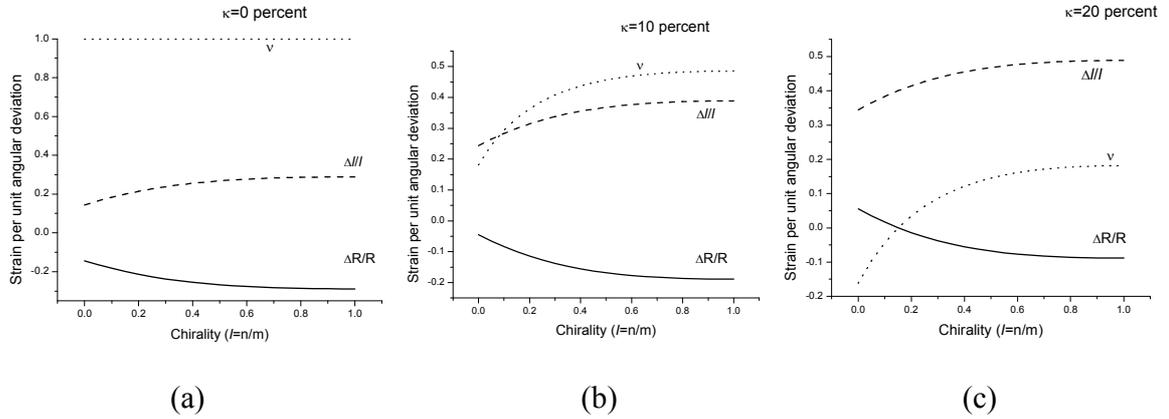

(a)            (b)            (c)

Fig. 5. Poisson ratio of SWNT for all chiralities. The results obtained for lateral ($\Delta R/R$) (Eq. 15) and axial ($\Delta l/l$) strains (Eq. 19) per unit $d\theta$ are also shown at fixed bond lengths (a), at 10% flexibility of bond lengths (b) and at 20% flexibility of bond lengths (c). The flexibility $\kappa$ is defined as Eq. 20.

It is interesting to note that as the flexibility of bond lengths is allowed to increase, there is compensation for lateral strain. The bond angle flexibility turns it one way, whereas the bond length flexibility turns it the other way. Interesting observations are possible, especially for low chirality tubes, where the Poisson ratio for $\kappa \sim 20\%$, can even be negative, and increases with increase in chirality. Some values of Poisson ratio are given in Table 1.

Table 1. Poisson ratio $\nu$ for SWNT of various chiralities, $I$ at some percent bond flexibility of the bond angle.

| S.No. | Chirality $I$=n/m | Poisson Ratio $\nu$ | | |
|---|---|---|---|---|
| | | $\kappa$=0% | $\kappa$=10% | $\kappa$=20% |
| 1 | 0 | 1 | 0.18 | -0.16 |
| 2 | 0.25 | 1 | 0.39 | 0.06 |
| 3 | 0.5 | 1 | 0.46 | 0.15 |
| 4 | 0.75 | 1 | 0.48 | 0.17 |
| 5 | 1 | 1 | 0.49 | 0.18 |

On comparison with the results for Poisson ratio from previous calculations[1], these vary between 0.19 and .22 for some tubes of various chosen diameters. This strengthen our generalized observations.

## 5. Concluding Remarks

We show that Poisson ratio of carbon nanotubes can be determined on the basis of geometrical structures alone, under the assumption that bond lengths forming the carbon nanotubes are rigid. In that case, the details of molecular potential is unnecessary for determination of Poisson ratio. Interesting numerical results have been obtained and it seems very natural for any cage like structures to give similar results. When both bond length and bond angle are assumed flexible, there is quite a lot of variation in the values of Poisson ratio because of chirality. The results for Poisson ratio do depend on chirality but not on the radius of same chirality. This is an important observation.

It will be interesting to measure the Poisson ratio of various chirality tubes.

**References:**

1. A. Sears and R.C. Batra, Phys. Rev. B **69**,235406(2004), **and references cited therein.**
2. S. Reich, C. Thomsen and P. Ordejon, Phys. Rev. B **65**,153407(2002).
3. J.P. Lu, Phys. Rev. Lett. 79, 1297-1300(1997).
4. C. Goze, L. Vaccarini, L. Henrard, P. Bernier E. Hernandez, and A. Rubio *Synthetic Metals. **103** (1999) 2500*
5. L. Vaccarini, C. Goze, L. Henrard, E. Hernandez, P. Bernier, and A. Rubio *Carbon. 38 (2000) 1681*
6. G. Dereli and C. Ozdogan arXiv:cond-mat/0303391 v1 19 Mar 2003
7. Shuchi Gupta, K. Dharamvir and V.K. Jindal (Submitted to PRB, 2005)